\documentclass[final]{siamltex}

\usepackage[latin1]{inputenc}
\usepackage{amsmath}
\usepackage{amssymb}
\usepackage{graphicx}

\newcommand{\m}[1]{\boldsymbol{#1}}
\newcommand{\T}{\text{T}}
\newcommand{\matrise}[1]{\begin{bmatrix} #1 \end{bmatrix}}

\newcommand{\diff}{\text{d}}
\newcommand{\pot}[2]{#1 \cdot 10^{#2}}

\title{Inverse scattering in multimode structures}

\author{Ole Henrik Waagaard\thanks{Optoplan AS, NO-7448 Trondheim,
    Norway, ({\tt ole-henrik.waagaard@
      eu.weatherford.com}).}  \and J. Skaar\thanks{Department of
    Electronics and Telecommunications, Norwegian University of
    Science and Technology, NO-7491 Trondheim, Norway.}}

\begin{document}

\maketitle

\begin{abstract}
  We consider the inverse scattering problem associated with any
  number of interacting modes in one-dimensional structures. The
  coupling between the modes is contradirectional in addition to
  codirectional, and may be distributed continuously or in discrete
  points. The local coupling as a function of position is obtained
  from reflection data using a layer-stripping type method, and the
  separate identification of the contradirectional and codirectional
  coupling is obtained using matrix factorization. Ambiguities are
  discussed in detail, and different {\it a priori} information that
  can resolve the ambiguities is suggested. The method is exemplified
  by applications to multimode optical waveguides with
  quasi-periodical perturbations.
\end{abstract}



\pagestyle{myheadings}
\thispagestyle{plain}
\markboth{O. H. WAAGAARD AND J. SKAAR}{MULTIMODE INVERSE SCATTERING}

\section{Introduction}
In waveguides that support several modes, scattering, or coupling
between the different modes, may appear due to different kinds of
perturbations. Possible perturbations are reflectors, gratings, bends,
tapering, and other kind of geometrical or material modulation along
the waveguide. The coupling may be both codirectional (coupling
between modes that propagate in the same direction) or
contradirectional (coupling between modes that propagate in opposite
directions). The direct scattering problem of computing the scattered
field when the probing waves and the scattering structure are known
has been extensively discussed in the literature
\cite{marcuse,snyder,kogelnikbook}. The inverse scattering problem
associated with two interacting modes is also well understood, and has
been treated in several contexts since the pioneering work by Gel'fand
and Levitan \cite{GLM}, Marchenko \cite{marchenko}, and Krein
\cite{krein}. In geophysics the so-called dynamic deconvolution or
layer-stripping (layer-peeling) methods emerged, for the
identification of layered-earth models from acoustic scattering data
\cite{pekeris, robinson, bardan, brucksteintrans, brucksteinrev}. More
recently the inverse scattering methods have been applied to the
design and characterization of optical devices involving two
interacting modes. Both contradirectional coupling and codirectional
coupling have been treated. Optical components based on
contradirectional coupling include thin-film filters and fiber Bragg
gratings \cite{songfbg, feced, skaarlp, rakesh, skaar_waagaard,
  rosenthal}, while codirectional coupling is present in e.g.
grating-assisted codirectional couplers and long-period gratings
\cite{jinguji,songlpg,fecedlpg,wanglpg,brenne}.  While the
inverse-scattering problem associated with two interacting modes is
well-known, the inverse-scattering problem of several, possibly
non-degenerate modes (i.e., with different propagation constants)
seems unsolved so far. Some work has been done in the case of 4
degenerate modes, that is, two polarization modes in each direction
\cite{sandel,waagaardpol}, and several degenerate modes with only
contradirectional coupling \cite{Aktosun}.

On the other hand, several methods for the inverse scattering of
acoustic or electromagnetic waves in two or three dimensions have
been reported. In particular, Yagle {\it et al.} have developed
layer-stripping methods for the multidimensional case
\cite{YagleL86,Yagle86,Yagle_multich}. By Fourier transforming the
problem with respect to the transversal coordinates, the
multidimensional problem may be regarded as one-dimensional with
several interacting modes.

In this paper we will extend these lines of thought to cover the
general inverse scattering problem associated with any number of
interacting modes in one-dimensional, reciprocal structures. In the
model (Section 2) both codirectional and contradirectional coupling
may be present simultaneously. We limit ourselves to the case where
the known probing waves and the scattered waves propagate in opposite
directions. In other words the scattered wave is considered as a
reflection from the unknown structure. A layer-stripping inverse
scattering algorithm is presented in Section 3. Ambiguities related to
the simultaneous presence of co- and contradirectional coupling are
discussed in detail. Possible {\it a priori} information that can
resolve these ambiguities will be suggested. The formalism is
particularly useful for the quasi-periodical case (Section 4), since
only the slowly varying envelope needs to be represented rather than
the structure itself, yielding an efficient algorithm. In Section 5,
the method is applied for the numerical reconstruction of a
quasi-periodical waveguide structure. Sections 4 and 5 are exemplified
by a multimode fiber Bragg grating; an optical fiber with
quasi-periodic refractive index perturbation along the fiber axis,
giving rise to both co- and contradirectional coupling. Finally,
analogies to the multidimensional case are discussed in Section 6.

\section{Continuous and discrete coupling model} 
\label{sec:model}

Consider a structure with $P$ modes propagating in each direction
along the $x$-axis. We visualize the $x$-axis as being directed to the
right, and say that the $+x$-direction is the forward direction. The
propagation constant of the $p$th mode is $\pm \beta_p$, i.e., the
$x$-dependence of the complex field associated with mode $p$ is
described by the factor $\exp(\pm i\beta_p x)$, where the upper
(lower) sign applies to forward (backward) propagating modes. Note
that the propagation constants of different modes may or may not be
different. The propagation constants are related to frequency through
the dispersion relation of the structure. The propagation constants
may be expressed $\beta_p=n_p\omega/c$, where $\omega$ is the angular
frequency, $c$ is some fixed reference velocity (common for all
modes), and $n_p$ accounts for the actual phase velocity. (However, in
some cases it may rather be convenient to express the propagation
constants in the form $n_p\omega/c-\pi/\Lambda$, where $\Lambda$ is a
constant, see Section \ref{sec:cont-coupl-struct}.) For
electromagnetic waves, it is natural to set $c$ equal to the vacuum
velocity, and consequently we will refer to $n_p$ as the effective
index associated with mode $p$. In principle, the effective indices
may be complex and dependent on frequency, meaning that modal loss and
dispersion are permitted in the model. However, the dispersion must be
limited by relativistic causality in the sense that any signal carried
by the modes travels no faster than the vacuum light velocity. Also,
the modal field profiles are assumed to have uniform phases such that
they can be written real.

Coupling may occur due to a continuous or discrete scattering
structure. In the first case, the field is assumed to be governed by
the coupled-mode equation
\begin{equation} 
\label{eq:CME}
\frac{\diff\mathbf E}{\diff x}=i\mathbf C\mathbf E,
\end{equation} 
where $\mathbf E$ is a column vector containing the $2P$ mode
amplitudes. In the absence of the scattering structure ($\mathbf C_{\m
  \sigma}=\mathbf C_{\m \kappa}=0$, see below), the first $P$ elements
are the mode amplitudes of the forward propagating modes (propagating
in the $+x$ direction) and the last $P$ elements are those of the
backward propagating modes. The coupling matrix $\mathbf C$ can be
decomposed into three contributions:
\begin{equation}
\label{eq:Ccontr}
\mathbf C=\mathbf D+\mathbf C_{\m \sigma}+\mathbf C_{\m \kappa}. 
\end{equation}
The contributions can be expressed as $2\times 2$ block matrices
consisting of $P\times P$ blocks:
\begin{subequations}
\label{eq:Cmatrs}
\begin{align}
  \label{eq:D}
  \mathbf D&=\matrise{\m \beta&\m 0\\\m 0&-\m \beta}, \\ 
  \label{eq:Cq}
  \mathbf C_{\m\kappa}&=\matrise{\m 0&\m \kappa\\ -\m \kappa^* &\m 0}, \\
  \label{eq:Cdc}
  \mathbf C_{\m\sigma} &=\matrise{\m\sigma&\m 0\\ \m 0&-\m\sigma^*},
\end{align}
\end{subequations}
where * denotes complex conjugate. The first term $\mathbf D$
describes the frequency dependence due to the propagation of the
different modes (``self-coupling''), and is independent on $x$; and
$\m\beta=\diag\{\beta_1, \beta_2,\ldots, \beta_P\}$. Only this term is
permitted to be lossy in the model ($\m\beta$ may be complex). In practice, we should require $\text{Im}\:\beta_p L\lesssim
  1$, where $L$ is the total length of the structure; otherwise the
  field at the far end of the structure may be close to zero (i.e.,
  the mode will be bound at the left interface to the structure, and
  very little reflection will originate from the far end.). The
second term $\mathbf C_{\m \kappa}$ describes the coupling between
counterpropagating modes, whereas the last term $\mathbf C_{\m\sigma}$
accounts for the coupling between copropagating modes. The coupling
coefficients $\m \kappa$ and $\m\sigma$ are dependent on $x$ but
assumed independent on frequency.  As will become clear shortly, the
above forms of $\mathbf C_{\m \kappa}$ and $\mathbf C_{\m\sigma}$ are
consequences of reciprocity and losslessness. It should be
  noted that in structures such as long-period gratings, where the
  coupling is purely codirectional, the coupling is described by $\m
  \kappa=\m 0$ and a $\m\sigma$ with non-zero off-diagonal elements.
  The conventional way of describing such structures would be only to
  consider the upper-left $P\times P$ block of $\mathbf C$. The
  layer-stripping method in Sec. \ref{sec:lp} cannot be used to
  reconstruct such structures since the reflection response is zero.

The coupling region in the waveguide is discretized into $N$ layers,
each of thickness $\Delta x=L/N$. If $N$ is sufficiently large so that
the matrices in \eqref{eq:Cmatrs} can be treated as constants in each
layer, we can solve \eqref{eq:CME}:
\begin{equation}
\label{eq:CMEsolution}
\mathbf E(x_j+\Delta x)=\exp{(i\mathbf C \Delta x)}\mathbf E(x_j),
\quad x_j=j\Delta x. 
\end{equation}
This transfer matrix relation can be used to propagate the fields
through the piecewise uniform structure. With the help of the connection
between the transfer matrix and the scattering matrix (Appendix
\ref{sec:gencomp}) we can find the reflection and transmission response
from the total transfer matrix, obtained as a product of the transfer
matrices $\exp{(i\mathbf C \Delta x)}$ of each layer (direct
scattering). 

While direct scattering is achieved straightforwardly using the
piecewise-uniform discretization, for inverse scattering it is
convenient to push the discretization further, to identify the
different contributions to the transfer matrix $\exp{(i\mathbf C
  \Delta x)}$. To first order in $\Delta x$, we have $\exp(i\mathbf
C\Delta x)=\exp(i\mathbf D\Delta x)\exp(i\mathbf C_{\m \kappa}\Delta
x)\exp(i\mathbf C_{\m\sigma} \Delta x)$. For a continuous structure of
finite thickness, the bandwidth where the reflection spectrum is
significantly different from zero is finite. Thus we need only be
concerned with frequencies satisfying $|\omega|\leq\omega_{\text{b}}$
for some positive constant $\omega_{\text{b}}$. Note that
  this model may give entirely incorrect results for
  $|\omega|>\omega_{\text{b}}$. For instance, if $P=1$ the reflection
  spectrum calculated with the discrete model will be periodic with
  period $\pi c/(n_1\Delta x)$, while the spectrum associated with a
  continuous structure tends to zero for large frequencies. For
inverse scattering, the reflection spectrum and therefore
$\omega_{\text{b}}$ are known. Therefore, provided $\Delta x$ is
chosen sufficiently small we can approximate each layer by a cascade
of three sections: a section with codirectional coupling, a section
with contradirectional coupling, and time-delay section.  The physical implication of this factorization is that the
  mode-coupling appears in a discrete point within the layer rather
  than distributed along the whole layer. The contradirectional
  section may therefore be pictured as a discrete reflector. The
transfer matrix of the $j$th layer becomes
\begin{equation}
\label{eq:51mm}
  \mathbf T_j
  =\mathbf T_{\m Z}\mathbf T_{\m\rho_j}\mathbf T_{\m\Phi_{j}},
\end{equation}
where
\begin{subequations}
\label{eq:Tmatrs}
\begin{align}
  \label{eq:TZ1}
  \mathbf T_{\m Z}&\equiv \exp(i\mathbf D\Delta x)=
  \matrise{\m Z^{-1}&\m 0\\\m 0 &\m Z},
  \quad \m Z^{-1}=\exp(i \m\beta \Delta x),\\
  \label{eq:Trho1}
  \mathbf T_{\m \rho_j}&\equiv\exp(i\mathbf C_{\m\kappa}\Delta x)
  =\matrise{\m t_j^{-1*} & -\m t_j^{-1*}\m\rho_j^* \\ 
    -\m t_j^{-1}\m\rho_j& \m  t_j^{-1}},
  \quad \begin{array}{l}
    \m\rho_j=i\tanh[(\m\kappa^*\m\kappa)^{1/2}\Delta x]
    (\m\kappa^*\m\kappa)^{-1/2}\m\kappa^*,\\ 
   \m t_j =\cosh[(\m\kappa^*\m\kappa)^{1/2}\Delta x]^{-1},   
   \end{array}\\
  \label{eq:TPhi1}
  \mathbf T_{\m \Phi_{j}}&\equiv\exp(i\mathbf C_{\m\sigma} \Delta x)
  =\matrise{\m \Phi_{j} &\m 0\\ \m 0&\m \Phi_{j}^*}, \quad \m \Phi_{j}=\exp(i\m\sigma\Delta x).
\end{align}
\end{subequations}
The form of the matrix in \eqref{eq:Trho1} may for example be verified
by evaluating the power series expansion of the matrix exponential. In
principle, it suffices to express \eqref{eq:Tmatrs} to first order in
$\Delta x$; however, the exact form is kept to emphasize the
properties of each of the three sections, to ensure that
  each section is lossless regardless of the value of $\Delta x$, and
  to retain the correspondence to the discrete case (below).

We are now in the position that we can argue for the forms of the
coupling matrices \eqref{eq:Cmatrs}. Note that while we have permitted
loss in the propagation section $\m Z^{-1}$, the coupling sections are
assumed lossless. Since the coupling sections also are assumed to be
reciprocal, their transfer matrices satisfy \eqref{eq:49} and
\eqref{eq:50} (Appendix \ref{sec:gencomp}). Allowing a more general
$\mathbf C_{\m\kappa}$ by substituting $\m\kappa^*\to \m\kappa_{21}$
in the (2,1) block, and expanding $\exp(i\mathbf C_{\m\kappa}\Delta
x)$ to first order in $\Delta x$, the lossless and reciprocity
conditions give $\m\kappa_{12}=-\m\kappa^*$ and dictate $\m\kappa$ to
be symmetric.  Similarly, we can derive the form of $\mathbf
C_{\m\sigma}$ and establish that $\m\Phi_{j}$ must be unitary, i.e.,
$\m\sigma$ is hermitian.

From the discussion above, each layer is characterized by a unitary
codirectional coupling matrix $\m\Phi_{j}$ and a discrete reflector.
Let superscript T denote transpose and let $\|\cdot\|$ be the usual
matrix 2-norm. The discrete reflector satisfies $\m\rho_j=\m\rho_j^\T$
and $\|\m\rho_j\|< 1$, and has an associated, positive definite
transmission matrix $\m t_j$ with $\m t_j^2=\m I-\m \rho_j\m
\rho_j^*$. 

So far we have considered a continuous scattering structure, and
discretized it into a cascade of codirectional coupling, reflection,
and pure propagation. Obviously, we can also describe discrete
coupling directly. The most general, lossless, reciprocal coupling
element can be described as a discrete reflector sandwiched between
two codirectional coupling sections (Appendix \ref{sec:gencomp}).
Compared to our discrete model above, there is an extra codirectional
coupling section on the right-hand side of the reflector. In the
special case where all modes have equal effective index, $\m
Z^{-1}\propto \m I$, this coupling section commutes with the delay
section, and as a result it can be absorbed into the next, adjacent
layer on the right-hand side. However, in the general case this extra
coupling section does not commute with the delay section and cannot be
ignored.  For inverse scattering, this coupling section should
therefore not be present since otherwise, it would not be possible to
  determine the transmission through the layer uniquely from the
  reflection. Under this assumption, $\m t_j$ is positive semidefinite, and
  uniquely determined by $\m t_j^2=\m I-\m \rho_j\m\rho_j^*$. We restrict ourselves to reflectors that satisfy
  $\|\m\rho_j\|< 1$; otherwise the reflector will mask the later part
  of the structure such that the inverse scattering procedure will not
  be possible.  Also, with two or more layers with $\|\m\rho_j\|=1$,
  the structure may behave as an ideal resonator with bound modes.


Writing out the transfer matrix \eqref{eq:51mm} of each layer, we obtain
\begin{equation}
  \label{eq:51m}
  \mathbf T_j= 
  \matrise{ \m Z^{-1}\m t_j^{-1*}\m \Phi_{j} 
    & -\m Z^{-1}
    \m t_j^{-1*}\m\rho_j^*\m\Phi_j^*\\   
    -\m Z\m t_j^{-1}\m\rho_j\m \Phi_{j} & 
    \m Z \m t_j^{-1}\m \Phi_{j}^*}
  =\matrise{\m Z^{-1} \m K_j &\m 0 \\ \m 0 & \m Z\m K_j^* }
  \matrise{\m I & -\m \Upsilon_j^* \\ -\m \Upsilon_j & \m I}, 
\end{equation}
where $\m\Upsilon_j=\m\Phi_{j}^\T\m\rho_j\m\Phi_{j}$ and $\m K_j=\m
t_j^{-1*}\m\Phi_{j}$. The transfer matrix can be converted into a
scattering matrix (Appendix \ref{sec:gencomp}):
\begin{equation}
  \label{eq:46}
\mathbf S_j=\matrise{\m \Phi_{j}^\T \m \rho_j \m \Phi_{j} 
    & \m\Phi_{j}^\T \m t_j \m Z^{-1} \\  
    \m Z^{-1} \m t_j^* \m\Phi_{j} &
    - \m Z^{-1}\m t_j^{-1*}\m\rho_j^*\m t_j \m Z^{-1}}.
\end{equation}
Thus, $\m \Upsilon_j$ represents the reflection response from the left of layer $j$.

The combined transfer matrix describing the total structure with $N$
layers is given by
\begin{equation}
  \label{eq:59m}
  {\mathbf T}= {\mathbf T}_{N-1} {\mathbf T}_{N-2} \cdots {\mathbf
  T}_{1}{\mathbf T}_{0}.
\end{equation}
From this matrix we can determine the reflection and transmission
response using \eqref{eq:11}. For example, the reflection response
from the left is
\begin{equation}
\m R(\omega)\equiv\m S_{11}=-\m T_{22}^{-1} \m T_{21},
\end{equation}
where $\m T_{kl}$ are the $P\times P$ blocks in $\mathbf T$. Assuming
$\|\m\rho_j\|<1$ for all $j$, it can be proven by induction that $\m
T_{22}$ is invertible on and above the real frequency axis in the
complex $\omega$-plane, for any number of layers. Physically this is
obvious since $\m T_{22}^{-1}$ is the transmission response from the
right, and therefore it must exist and be causal and stable.

Reciprocity \eqref{eq:14} gives $\m R(\omega)=\m R(\omega)^\T$. Using
$\|\m\rho_j\|<1$ for all $j$, it can be shown by induction that $\|\m
R(\omega)\|<1$ for a passive structure (a passive structure is
characterized by $\text{Im}\,\beta_p\geq 0$ for all $p$). By causality
the reflection response can be written in the form
\begin{equation}
\label{eq:Rh2}
\m R(\omega)=\int_{0}^{\infty}\m h(t)\exp(i\omega t)\diff t,
\end{equation}
where $\m h(t)$ is called the time-domain impulse response. 

When the modes are nondispersive, i.e., $\m\beta$ is linearly related
to frequency, $\m h(t)$ equals a train of non-equally spaced,
weighted delta pulses:
\begin{equation}
  \label{eq:defg}
  \m h(t)=\sum_{k=0}^\infty \m h^k\delta(t-t^k).
\end{equation}
Here $\m h^k$ and $t^k$ are the weight and arrival time of the $k$th
pulse, respectively. Substituting \eqref{eq:defg} into \eqref{eq:Rh2}
gives
\begin{equation}
\label{eq:Rh1}
\m R(\omega)=\sum_{k=0}^\infty \m h^k\exp(i\omega t^k).
\end{equation}
The weights $\m h^k$ can in principle be calculated from $\m
R(\omega)$ using an inverse transform of the form
\begin{equation}
\label{eq:iF}
  \m h^k= \lim_{\omega_{\max}\to\infty}\frac{1}{2\omega_{\max}}
  \int_{-\omega_{\max}}^{\omega_{\max}} \m R(\omega)\exp(-i\omega t^k)\diff\omega.
\end{equation}
The arrival times are determined by the delay from a layer to the next
of each mode. Let $\Delta t_p$ be the delay of mode $p$ through a
single layer. A delta pulse at $t=0$ is incident to the structure on
the left-hand side. Consider the reflection from the different layers,
as seen from left-hand side of the structure. From layer 0, the
arrival times in all modes will be zero. An impulse in mode $p$
reflected from layer 1 into mode $q$, will arrive at $\Delta
t_p+\Delta t_q$. Thus, considering layer 1, the arrival times are any
combinations of two unit delays $\Delta t_p$. Considering layer 2, the
arrival times are any combinations of four unit delays, and so forth.

When the modes are dispersive, the impulse response is no longer a
train of delta functions. Nevertheless, for $t=0$ it can still be
written as $\m h^0\delta(t)$, and the weight $\m h^0$ can be found
from \eqref{eq:iF}.

Eq. \eqref{eq:Rh1} clearly demonstrate that, in principle, for a
discrete structure the reflection response $\m R(\omega)$ does not
approach zero for large frequencies. Only in the special case where
the modal effective indices are rational numbers with common
denominator, the reflection spectrum is periodic. Fortunately, in
practice, it is not necessary to represent the entire bandwidth to
enable inverse scattering for a discrete structure. As shown in the
next section, what is needed in the layer-stripping algorithm is the
zeroth point of the impulse response, at time $t=0$. Since the next
nonzero value is for $t=2\min_p\Delta t_p$ \footnote{For simplicity it
  is assumed a nondispersive structure.}, the zeroth point is computed
accurately provided the represented bandwidth $\omega_{\max}$
satisfies $\omega_{\max}\gg1/\min_p\Delta t_p$. Then, if the true
reflection spectrum is multiplied by a smooth window function
$W(\omega)$ that goes to zero at $\omega=\pm\omega_{\max}$, the
inverse Fourier transform evaluated around zero is approximately
$w(t)\m h^0$, where $w(t)$ is the inverse Fourier transform of
$W(\omega)$. Since $w(0)\m
h^0\approx\frac{1}{2\pi}\int_{-\omega_{\max}}^{\omega_{\max}}
W(\omega)\m R(\omega)\diff\omega$, we can find $\m h^0$ from a
measurement of $\m R(\omega)$ in the bandwidth
$(-\omega_{\max},\omega_{\max})$:
\begin{equation}
\label{eq:iFw}
\m h^0\approx\frac{\int_{-\omega_{\max}}^{\omega_{\max}} W(\omega)\m R(\omega)\diff\omega}
{\int_{-\omega_{\max}}^{\omega_{\max}} W(\omega)\diff\omega}.
\end{equation}

In many practical cases, the structure to be reconstructed is
quasi-sinusoidal. More generally, the structure is often
quasi-periodic, and e.g. the first ``Fourier component'' is to be
reconstructed. In such cases, one can define modal field envelopes
which vary slowly with respect to $x$ (compared to a wavelength).
Similarly, one can extract slowly varying coupling coefficient
envelopes. As a result, all quantities in \eqref{eq:CME} vary slowly
with $x$. The relevant bandwidth in \eqref{eq:iFw} will then be
centered about a chosen ``design frequency'' rather than zero. The
main advantage of this procedure is that it leads to considerably less
requirements on the spatial resolution, and as a result efficient
inverse scattering. This modification to the model is detailed in
Section \ref{sec:cont-coupl-struct}.

\section{Layer-stripping method}
\label{sec:lp}
The inverse scattering problem can now be stated as follows: Given a
structure consisting of $N+1$ layers. Each layer consists of three
sections (sublayers), the first ($\m\Phi_j$) responsible for coupling
between copropagating modes, the second ($\m\rho_j$) responsible for
coupling between counterpropagating modes, and the third a pure
propagating section ($\m Z^{-1}$). The propagation constants of the
involved modes are known and specified in terms of $\m Z^{-1}$.
\footnote{The effective indices may contain small, real, unknown parts
  $\Delta n_p$, i.e., $n_p=n_{p,\text{known}}+\Delta n_p$, where
  $n_{p,\text{known}}$ are known. Provided $\Delta n_p$ is
  sufficiently small, the variation of the associated phase factor
  $\exp(i\omega\Delta n_p\Delta x/c)$ may be small over the relevant
  bandwidth. In such cases the unknown parts can be absorbed into the
  $\m\Phi_j$'s.} From a set of excitation-response pairs (that is, $\m
R(\omega)$), we want to reconstruct $\m\rho_j$ and $\m\Phi_j$ for all
$j$.

The structure itself and the medium to the right are assumed to be at
rest at time $t=0$. For incident waves from the left, the reflection
response from the structure is described by the matrix $\m R(\omega)$
of dimension $P \times P$. This matrix can be viewed as the operator
which takes the excitation field vector to the reflected field vector.
Its columns can be interpreted as the responses for orthonormal
excitation basis vectors $\m e_1,\m e_2,\dotsc,\m e_P$, respectively.
Here $\m e_p$ has only one nonzero element (equal to unity) at
position $p$. Similarly, we can define the forward ($\m u_j(\omega)$)
and backward ($\m v_j(\omega)$) propagating field matrices as $P\times
P$ matrices where the columns are the fields for orthonormal
excitations $\m e_1,\m e_2,\dotsc,\m e_P$.  A subscript $j$ is
specified to emphasize that $\m u_j(\omega)$ and $\m v_j(\omega)$ are
the fields at the beginning (left-hand side) of layer $j$.  The field
matrices of layer $j+1$ are related to the field matrices of layer $j$
by
\begin{equation}
\label{eq:1}
\matrise{\m u_{j+1}(\omega) \\\m v_{j+1}(\omega)}=
\mathbf T_j \matrise{\m u_{j}(\omega) \\ \m v_{j}(\omega)},
\end{equation}
where $\mathbf T_j$ is given by \eqref{eq:51m}.

The layer-stripping algorithm is based on the simple fact that the
leading edge of the impulse response is independent on later parts
of the structure due to causality. Hence, one can identify the first
layer of the structure, and subsequently remove its effect using the associated transfer matrix.

For layer 0, we initialize $\m u_0(\omega)=\m I$ and $\m
v_0(\omega)=\m R(\omega)$. We define a local reflection spectrum $\m
R_j(\omega)=\m v_j(\omega)\m u_j(\omega)^{-1}$ and the associated
impulse response $\m h_j(t)$ as the response of the structure after
removing the first $j-1$ layers. Similarly to the impulse response of
the entire structure, $\m h_j(t)$ contains an isolated delta function
at $t=0$. Due to causality, this pulse is equal to the reflection from
the zeroth layer alone. Denoting the weight of this pulse $\m h^0_j$,
we find from \eqref{eq:46} that
\begin{equation}
\m h^0_j=\m\Upsilon_j\equiv \m\Phi_{j}^\T\m\rho_j\m\Phi_{j}.
\end{equation}
Note that $\m R_j(\omega)$ is symmetric for all $\omega$ as a result
of reciprocity; thus $\m h^0_j$ is symmetric as well. Writing out
\eqref{eq:1} and \eqref{eq:51m}, and substituting $\m v_j(\omega)=\m
R_j(\omega)\m u_j(\omega)$, we obtain
\begin{subequations}
\label{eq:3}
  \begin{align}
    \m u_{j+1}(\omega)&=\m Z^{-1}\m K_j\left[\m I-\m \Upsilon_j^* \m R_j(\omega)\right]
    \m u_j(\omega),\\
    \m v_{j+1}(\omega) &=\m Z\m
    K^*_j\left[ \m R_j(\omega)-\m \Upsilon_j\right]\m u_j(\omega),
  \end{align}
\end{subequations}
and therefore
\begin{equation}
  \label{eq:shur}
  \m R_{j+1}(\omega)=\m Z\m K^*_j\left[ \m R_j(\omega)-\m \Upsilon_j\right]
  \left[\m I-\m \Upsilon_j^* \m R_j(\omega)\right]^{-1}\m K_j^{-1} \m Z.
\end{equation}
Provided $\m\Upsilon_j$ and $\m K_j$ are known, \eqref{eq:shur} shows
that the local reflection spectrum of layer $j+1$ can be calculated
directly from the local reflection spectrum of layer $j$ without
calculating the fields $\m u_{j+1}$ and $\m v_{j+1}$. Note the
similarity to the Schur formula used in scalar layer-stripping
\cite{brucksteintrans}.

To characterize layer $j$ completely, and to identify $\m K_j$, we
must determine $\m\rho_j$ and $\m\Phi_j$. By counting the available
degrees of freedom (in $\m\Upsilon_j$), we immediately find that this
cannot be done uniquely. It is therefore necessary to use {\it a
  priori} information on $\m\rho_j$ and/or $\m\Phi_j$. The available
information may vary from situation to situation. Here we will
consider the following situations, where $\m\rho_j$ and $\m\Phi_j$ can
be found using the methods in the Appendices \ref{sec:svdsymm} and
\ref{sec:UPQfact}.

\renewcommand{\labelenumi}{\alph{enumi})}
\begin{enumerate}
\item \label{item:gen} $\m\Phi_j=\m I$. In this case there is no
  codirectional coupling. The identification of the layer is now
  particularly simple, as $\m\rho_j=\m\Upsilon_j$ uniquely. Note that
  while there is no codirectional coupling, $\m\rho_j$ describes
  reflection from all modes into all modes. Thus the different modes
  may still interact.
  
\item \label{item:diag} $\m\rho_j$ is diagonal and nonnegative. Now
  $\m\rho_j$ is a simple partial reflector which only reflects light
  into the same mode as the incident field (no reflection into other
  modes). The coupling between different modes is instead described by
  $\m\Phi_j$.  Since $\m\Upsilon_j=\m\Phi_{j}^\T\m\rho_j\m\Phi_{j}$,
  $\m\rho_j$ is found uniquely as the singular value matrix associated
  with $\m\Upsilon_j$, up to reordering of the singular values. Once
  the order of the singular values has been established, the unitary
  $\m\Phi_j$ is found uniquely up to the sign of its rows, provided
  all singular values are distinct and nonzero (see Appendix
  \ref{sec:svdsymm}). When one or more singular values of
  $\m\Upsilon_j$ are zero, the corresponding row(s) of $\m\Phi_{j}$
  cannot be determined uniquely. More precisely, $\m\Phi_{j}$ is
  determined up to a premultiplicative unitary matrix $\m J$ operating
  on the associated mode(s). Physically, this is obvious since when a
  singular value is zero, the associated mode is not reflected from
  the layer. When two or more nonvanishing singular values are equal,
  $\m\Phi_j$ is determined up to a premultiplicative, real unitary $\m
  J$ operating on the associated modes. Physically, this means that
  these modes experience the same reflection and thus an arbitrary
  (real) ``rotation'' of the modes is not detected. In such cases, the
  unitary section $\m\Phi_{j}$, as determined by the method in
  Appendix \ref{sec:svdsymm}, does not necessarily correspond to the
  physical section. This error will propagate to the next layers
  according to \eqref{eq:shur}.
  
\item \label{item:PQ} $\m\Phi_j$ is symmetric and $\m\rho_j$ is real and positive
  semidefinite.  A special case in which there are only two degenerate
  modes in each direction is treated in \cite{waagaardpol}. The
  reflector matrix $\m\rho_j$ can be written $\m P_j^{\T}\m\Sigma_j \m
  P_j$, where $\m P_j$ is a real, special unitary matrix and
  $\m\Sigma_j$ is diagonal and nonnegative. Since
  $\m\Upsilon_j=\m\Phi_{j}^\T\m\rho_j\m\Phi_{j}=\m\Phi_{j}^\T\m
  P_j^{\T}\m\Sigma_j\m P_j\m\Phi_{j}$, we find $\m\Sigma_j$ and $\m
  P_j\m\Phi_{j}$ as in the previous case, with the identical
  ambiguity issues. The separate identification of $\m P_j$ and
  $\m\Phi_{j}$ is accomplished using the factorization method in
  Appendix \ref{sec:UPQfact}, with certain ambiguities related to the sign
  of the eigenvalues of $\m\Phi_j$.
\end{enumerate}

The ambiguities when determining $\m\Phi_j$ in situation b) are in
fact very similar to the well-known ambiguities in the scalar case
with a single mode in each direction. In the scalar case any $\pi$
phase-shift sections between the reflectors cannot be identified since
the associated round-trip phase accumulated to and from a reflector
becomes $2\pi$. In our multimode case, the sign of the rows of the
``phase-delay'' section ($\m\Phi_j$) between two reflectors cannot be
identified. Similarly, in the scalar case, any phase-shift section
preceeding a zero reflector cannot be determined uniquely. Instead it
is chosen arbitrarily (e.g. removed), and attributed to
the next layer with a nonzero reflector.

When the structure to be reconstructed is a discretized version of a
smooth structure, the smoothness can be used to resolve ambiguites.
First we consider situation b). For small $\Delta x$, $\m\Phi_j$ is
close to identity; thus the sign of the rows of $\m\Phi_j$ can be
determined uniquely. If $\m\rho_j$ has distinct eigenvalues, valid for
all $j$, the order of the eigenvalues of $\m\rho_j$ can be determined
from the order of the eigenvalues of $\m\rho_{j-1}$ using the
smoothness of $\m\kappa=\m\kappa(x)$. If there are equal eigenvalues
for a certain reflector $\m\rho_j$, or if $\m\rho_j$ is singular, the
ambiguities of $\m\Phi_j$ are characterized by the premultiplicative
$\m J$ matrix (Appendix \ref{sec:svdsymm}). In other words, the chosen
$\m\Phi_j$ is related to the corresponding true matrix
($\m\Phi_{j,\text{true}}$) by $\m\Phi_j=\m J\m\Phi_{j,\text{true}}$.
By choosing $\m J$ such that $\|\m\Phi_j-\m\Phi_{j-1}\|$ is minimum,
the resulting $\m J$ is close to identity (that is, $\|\m J-\m I\|
\leq 2\|\m\Phi_{j,\text{true}}-\m\Phi_{j-1}\|$). Since $\m t_j$ and
$\m Z^{-1}$ are close to identity as well, the order of three sections
$\m J$, $\m t_j$, and $\m Z^{-1}$ can be interchanged (see Section
\ref{sec:model}). Thus the error due to wrong choice of $\m\Phi_{j}$
can be absorbed into $\m\Phi_{j+1}$. More generally, provided only a
few neighboring layers have singular or degenerate $\m\rho_j$'s, only
the corresponding and following $\m\Phi_j$ sections may be determined
erroneously, and the determination of the later part of the structure
is (approximately) unaffected.

In situation c), the order of eigenvalues of $\m\rho_j$ can be
determined as in situation b). However, $\m P_j\m\Phi_j$ is not
necessarily close to identity. Nevertheless, the sign of its rows can
be determined from $\m P_{j-1}\m\Phi_{j-1}$ if $\m\kappa=\m\kappa(x)$
is sufficiently smooth. (Recall that $\m P_j\m\Phi_j$ is unitary,
which means that in each row there exists at least one element of
magnitude $\geq 1/\sqrt P$.) Finally, since $\m\Phi_j$ is close to
identity, its eigenvalues are close to unity. It follows that the
factorization of $\m P_j\m\Phi_j$ into $\m P_j$ and $\m\Phi_j$ is
unique (Appendix \ref{sec:UPQfact}).

From the discussion above, we summarize the layer-stripping algorithm,
analogously to the scalar version described in ref.
\cite{brucksteintrans,brucksteinrev}, that can be applied to identify a
structure supporting multiple modes:

\renewcommand{\labelenumi}{\arabic{enumi})}
\begin{enumerate}
\item
Initialize $j=0$. Set $\m R_j(\omega)=\m R(\omega)$.
\item \label{item:1} Compute the zeroth weight $\m h_j^0$ of the
  impulse response. In practice this is achieved by the substitutions
  $\m h^0\to\m h^0_j$ and $\m R(\omega)\to\m R_j(\omega)$ in
  \eqref{eq:iFw}.
\item Use a model-specific factorization of $\m
  h_j^0=\m\Phi_j^\T\m\rho_j\m\Phi_j$ to find $\m\Phi_{j}$ and
  $\m\rho_j$.
\item \label{item:4} Calculate $\m t_j=(\m
  I-\m\rho_j\m\rho_j^*)^{1/2}$ such that the associated eigenvalues
  are positive, and set $\m K_j=\m t_j^{-1}\m\Phi_{j}$.
\item Calculate the next, local reflection response $\m R_{j+1}(\omega)$
  using \eqref{eq:shur}.
\item If $j<N-1$, increase $j$ and return to \ref{item:1}.
\end{enumerate}

When the scattering structure is continuous, one can use the true
reflection spectrum as input to the layer-stripping algorithm, even
though the structure is modelled discrete. This can be justified as
follows: The layer thickness $\Delta x$ is chosen small such that the
first order approximations of $\exp(i\mathbf{C}_{\m\kappa}\Delta x)$
and $\exp(i\mathbf{C}_{\m\sigma}\Delta x)$ are accurate. (Thus an
upper bound on $\|\mathbf{C}_{\m\kappa}\|$ and
$\|\mathbf{C}_{\m\sigma}\|$ should be known {\it a priori}.) Let
$\omega\leq\omega_{\text{b}}$ be the bandwidth where the true
reflection spectrum is significantly different from zero. For
sufficiently small $\Delta x$, the first order approximation of
$\exp(i\mathbf{D}\Delta x)$ is valid, and the true reflection spectrum
is approximately equal to that of the corresponding discrete model in
the bandwidth $\omega\leq\omega_{\text{b}}$.  In the limit $t\to 0^+$,
the $(p,q)$ element of the impulse response of the continuous
structure can be calculated exactly from \eqref{eq:CME} using the Born
approximation, yielding
\begin{equation}
\label{eq:iFc}
h_{pq}(t=0^+)\equiv \frac{1}{2\pi}\lim_{t\to 0^+}\int_{-\infty}^{\infty} R_{pq}(\omega)\exp(-i\omega t)\diff\omega=i\kappa_{pq}^*(x=0^+)c/(n_p+n_q).
\end{equation}
Here $\kappa_{pq}(x=0^+)$ is the $(p,q)$ element of $\m\kappa(x)$ at
$x=0^+$. For practical computations, the integral in Eq.
\eqref{eq:iFc} must be truncated at $\pm\omega_{\text{b}}$; thus, to
find the leading edge of $h_{pq}(t)$, one can take $t=0$ in the
integral, and multiply the result by a factor of two. (Recall that by
causality $\lim_{t\to 0^-} h_{pq}(t)=0$.) Once $\m\kappa$ for the
zeroth layer is found, one can propagate the fields using
\eqref{eq:shur}. Since we have not identified the codirectional
coupling $\m\Phi_0$ of the zeroth layer, $\m\Phi_0$ is associated with
the next layer. Thus, after the zeroth layer has been stripped off,
the leading edge of the impulse response of the remaining structure
becomes
\begin{equation}
\label{eq:iFc1}
\m\Phi_0^\T \left[i\kappa_{pq}^*(x=\Delta x^+)c/(n_p+n_q)\right] \m\Phi_0,
\end{equation}
where the square bracket denotes a matrix formed by the elements
inside. The identification of $\m\Phi_0$ and
$\left[i\kappa_{pq}^*/(n_p+n_q)\right]$ can now be accomplished using
the factorization methods described above. The algorithm continues in
the same way, until finally the bandwidth of the reflection spectrum
of the remaining structure exceeds $\omega_\text{b}$. This remaining
part of the structure can be made arbitrarily thin by choosing a
sufficiently small $\Delta x$.

The difference between the latter ``quasi-continuous'' formulation and
the discrete algorithm is essentially the factor $n_p+n_q$, and the
method for evaluating the leading edge or first point of the impulse
response. When the effective indices can be approximated by some
number $n_0$ for all $p$, $n_p\approx n_0$, one can in fact use the
discrete algorithm directly: A periodic extension of the true
reflection spectrum outside a principal bandwidth
$[-\omega_{\text{max}},\omega_{\text{max}}]$ corresponds then to a
discrete model with $\Delta x=\pi c/(2n_0\omega_{\text{max}})$. The
first point of the impulse response is calculated by \eqref{eq:iFw}
using a rectangular window function $W(\omega)$. For a broad class of
waveguides of practical interest, the effective indices are similar
(see Section 4). While the phase relation between the modes, as
described by $\m Z^{-1}$, may still result in a nontrivial multimode
coupling, the discrete algorithm gives accurate results. The errors
due to this periodic spectrum approximation can be corrected to some
extent by including the factor $(n_p+n_q)/(2n_0)$ in the elements on
the right-hand side of \eqref{eq:iFw}. This can be justified e.g.
using the Born approximation.

\section{Quasi-sinusoidal coupling structures}
\label{sec:cont-coupl-struct}

Continuous coupling in acoustical, radio frequency, or optical
waveguides may be obtained by perturbation of the effective indices
$n_p$ associated with each mode. This can be achieved by modulation of
the wall profile or waveguide medium properties. As a concrete
example, we will discuss fiber Bragg gratings \cite{hillrev}, which
have attracted large interest recently due to their applications in
fiber optical communications and sensors. A fiber grating is formed in
an optical fiber by modulating the refractive index of the core
periodically or quasi-periodically. The main peak of the reflection
spectrum appears for the frequency where the reflection from a crest
in the index modulation is in phase with the next
reflection. Permanent gratings are fabricated by UV-illumination. In
fibers doped with certain dopants such as germanium, the
UV-illumination will permanently rise the refractive index of the
core. Advanced fabrication methods have made it possible to
manufacture complex gratings with varying index modulation amplitude
and period. The layer-stripping algorithm is the most widely used
method for designing the index profile to obtain a given reflection
spectrum \cite{feced,skaarlp,skaar_waagaard}.

In most cases, the fiber grating is formed in a single-mode fiber, and
coupling is only considered between the forward-propagating and
backward-propagating fundamental mode. The field matrices $\m
u_j(\omega)$ and $\m v_j(\omega)$ are then scalar functions. However,
in some cases it is not sufficient to consider only one
forward-propagating mode and one backward-propagating mode.  For
instance, a single mode fiber is always slightly birefringent, and the
photosensitivity can be polarization-dependent
\cite{hill_bilodeau}. In this case, two forward-propagating and two
backward-propagating polarization modes must be considered. An inverse
scattering algorithm that takes into account polarization mode
coupling is described in \cite{waagaardpol}. The coupling between the
two polarization modes are described by Jones matrices
\cite{jones}. Both polarization modes have approximately the same
effective index, so $\m Z^{-1}=\exp(i\beta\Delta x)\m I$, where the
common propagation constant $\beta$ is scalar.

In a multi-mode fiber, the modulation of the refractive index may
result in coupling between the fundamental mode and other modes. Each
mode has a transversal field profile $\Psi_p(r,\phi)$ which is a
solution to the scalar wave equation in polar coordinates $r$ and
$\phi$ \cite{snyder}\footnote{To find the exact electromagnetic modes,
the vector wave equation must be solved. However, for weakly guiding
waveguides (waveguides with small difference between the refractive
index of the core and the cladding), the scalar wave equation can be
used. This is the case for most conventional fibers.}:
\begin{equation}
  \label{eq:17}
  \left\{\nabla_{\text{t}}^2+k^2(\bar{n}^2(r)-n_p^2)\right\}\Psi_p(r,\phi)
  =0.
\end{equation} 
Here $\bar n(r)$ is the unperturbed, refractive index
profile of the fiber, which is assumed to be real, $\nabla_{\text{t}}$
is the transversal nabla operator, and $k=\omega/c$. The field
$\Psi_p(r,\phi)$ and its first derivatives are continuous. For bound
modes, the fields are real and orthonormal such that
$\int_{A_\infty}\Psi_p(r,\phi)\Psi_q(r,\phi)\diff A=\delta(p-q)$,
where $\delta(p-q)$ denotes the Kronecker delta, and $A_\infty$ is the
entire transversal plane. The effective indices $n_p$ are eigenvalue
solutions to \eqref{eq:17}. A mode $p$ is bound when
$n_\text{cl}<n_p\le n_\text{co}$, where $n_\text{co}$ and
$n_\text{cl}$ are the refractive indices of the fiber core and
cladding, respectively. Ignoring radiation modes, which in the
vicinity of the core decay rapidly away from the excitation source,
the total electric field $E(r,\phi,x)$ can be written as a
superposition of forward- and backward-propagating bound modes:
\begin{equation}
  \label{eq:71}
  E(r,\phi,x)=\sum_{p=1}^{P}(b_p^+(x)+b_{p}^-(x))\Psi_p(r,\phi).
\end{equation}
 Here $b_{p}^\pm(x)$ contain all $x$-dependence
including the harmonic propagation factor $\exp(\pm i \beta_px)$,
where $\beta_p=kn_p$.

Coupling between the modes originates from longitudinal modulation of
the refractive index. Let the refractive index be perturbed
quasi-periodically with a spatial period $\Lambda$,
\begin{equation}
  \label{eq:70} n(r,\phi,x)=\bar{n}(r)+\Delta n_\text{ac}(r,\phi,x)
  \cos\left(\frac{2\pi}{\Lambda}x+\theta(x)\right) +\Delta
  n_\text{dc}(r,\phi,x),
\end{equation} where $\Delta n_\text{ac}(r,\phi,x)$, $\Delta
n_\text{dc}(r,\phi,x)$, and $\theta(x)$ are slowly varying with $x$
over a distance $\Lambda$. We assume that $\Delta
n_\text{ac}(r,\phi,x)\ll \bar{n}$, and $\Delta
n_\text{dc}(r,\phi,x)\ll \bar{n}$, which is the case for practical
fiber gratings. The total electric field must satisfy the scalar
wave-equation for the perturbed fiber, i.e.,
\begin{equation}
\label{eq:72} \left\{\nabla_{\text{t}}^2+\frac{\partial^2}{\partial
  x^2}+k^2n^2(r,\phi,x)\right\}E(r,\phi,x)=0.
\end{equation} We now substitute \eqref{eq:71} into \eqref{eq:72},
take \eqref{eq:17} into account, and multiply the resulting equation
by $\Psi_q(r,\phi)$. By integration over the entire transversal plane,
and recalling that the modes are orthonormal, the resulting set of
second order differential equations can be decomposed into first order
coupled mode equations \cite{snyder},
\begin{subequations}
\label{eq:75}
  \begin{align}
    \label{eq:73} \frac{\diff b_p^+(x)}{\diff x}-i\beta_p
    b_p^+(x)&=i\sum_{q=1}^{P}\mathcal C_{pq}(x)(b_q^+(x)+b_{q}^-(x)),
    \\
    \label{eq:77} \frac{\diff b_{p}^-(x)}{\diff x}+i\beta_p
    b_{p}^-(x)&=-i\sum_{q=1}^{P}\mathcal
    C_{pq}(x)(b_q^+(x)+b_{q}^-(x)),
  \end{align}
\end{subequations} where
\begin{equation}
  \label{eq:74}
  \mathcal C_{pq}(x)=\frac{k}{2n_p}
    \int_{A_\infty}(n^2(r,\phi,x)-\bar{n}^2(r))\Psi_p(r,\phi)\Psi_q(r,\phi)\diff
    A.
\end{equation} Note that the frequency-dependence of \eqref{eq:74} can
be ignored in practice, since the normalized bandwidth of interest is
usually much less than unity, and the field profiles and effective
indices are approximately constant in this bandwidth. Also note that
since the fiber is assumed to be weakly guiding, $n_p$ can be set
equal to $n_{\text{co}}$; thus $\mathcal C_{pq}=\mathcal C_{qp}$.

In the case of a quasi-periodic structure it is natural to write the
coupling coefficient as a quasi-Fourier series:
\begin{equation}
\label{eq:quasisin}
\begin{split}
\mathcal C_{pq}(x)&=\sigma_{pq}(x) +
    \kappa_{pq}(x)\exp\left(i\frac{2\pi}{\Lambda}x\right)
    +\kappa^*_{pq}(x)\exp\left(-i\frac{2\pi}{\Lambda}x\right) \\
    &+\sum_{|m|\geq 2}\kappa^{(m)}_{pq}(x)\exp\left(i\frac{2\pi
    m}{\Lambda}x\right),
\end{split}
\end{equation}
where the ``Fourier coefficients'' $\kappa_{pq}(x)$,
$\sigma_{pq}(x)$, and $\kappa^{(m)}_{pq}(x)$ are slowly varying over a
period $\Lambda$. For a fiber grating the index modulation
$n(r,\phi,x)-\bar{n}(r)$ is given by \eqref{eq:70} and is small
compared to $\bar{n}(r)$, so the zeroth and first order Fourier
components dominate. Note that $\arg \{\kappa_{pq}(x)\}=\theta(x)$.

The field amplitudes $b_{p}^\pm(x)$ vary rapidly; it is therefore
convenient to introduce the slowly varying field envelopes $u_p(x)$
and $v_p(x)$ by setting
\begin{subequations}
  \label{eq:69}
  \begin{align}
    \label{eq:76}
    b_p^+(x)&=i^{1/2}u_p(x)\exp\left(i\frac{\pi}{\Lambda}x\right)\exp\left(i\frac{\theta(x)}{2}\right),\\
    \label{eq:78}
    b_{p}^-(x)&=i^{-1/2}v_p(x)\exp\left(-i\frac{\pi}{\Lambda}x\right)
    \exp\left(-i\frac{\theta(x)}{2}\right).
  \end{align}
\end{subequations}
Since an identical phase factor is removed from all
modes, the reflection response as calculated from $b_p^+$ and
$b_{q}^-$ will only differ from that calculated from $u_p$ and $v_q$
by a constant phase factor not dependent on $p$ and $q$. Inserting
\eqref{eq:quasisin} and \eqref{eq:69} into \eqref{eq:75}, and ignoring
rapidly oscillating terms (since they contribute little to $\diff
u_p/\diff x$ and $\diff v_p/\diff x$), we obtain an alternative set of
coupled-mode equations
\begin{subequations}
\label{eq:79}
  \begin{align}
    \label{eq:80} \frac{\diff u_p(x)}{\diff x}&=i\delta_p u_p(x)
    -\frac{i}{2}\frac{\diff\theta(x)}{\diff x}u_p(x)
     +i\sum_{q=1}^{P}\sigma_{pq}(x)u_{q}(x)
     +\sum_{q=1}^{P}|\kappa_{pq}(x)|v_q(x), \\
    \label{eq:81} \frac{\diff v_p(x)}{\diff x}&=-i\delta_p v_p(x)
    +\frac{i}{2}\frac{\diff \theta(x)}{\diff x}v_p(x)
     -i\sum_{q=1}^{P}\sigma_{pq}(x)v_{q}(x)
     +\sum_{q=1}^{P}|\kappa_{pq}(x)|u_q(x),
  \end{align}
\end{subequations} 
where $\delta_p=\beta_p-\pi/\Lambda=n_p\omega/c-\pi/\Lambda$ is the
wavenumber detuning of mode $p$. Thus, $-i|\kappa_{pq}(x)|$ is the
coupling coefficient between modes $p$ and $q$ propagating in opposite
directions, while $\sigma_{pq}(x)-\delta(p-q)(\diff\theta(x)/\diff
x)/2$ is the coupling coefficient between modes $p$ and $q$ in the
same direction.  With $\m E= [u_1, u_2, \ldots, u_P, v_1, v_2, \ldots,
v_P]^\T$ we find that \eqref{eq:79} coincides with \eqref{eq:CME},
where $\sigma_{pq}(x)-\delta(p-q)(\diff\theta(x)/\diff x)/2$ and
$-i|\kappa_{pq}(x)|$ are the $(p,q)$ elements of $\m\sigma$ and
$\m\kappa$, respectively, and $\delta_p$ are the diagonal elements of
$\m\beta$. Note that $\delta_p$ do not correspond to the actual
propagation constants but rather their detuning from $\pi/\Lambda$.
Approximating the effective indices by $n_{\text{co}}$, this means
that the bandwidth of interest is not centered about zero but rather
about the ``design frequency'' $\omega_0\equiv \pi
c/(n_{\text{co}}\Lambda)$. The frequency interval of integration in
\eqref{eq:iFw} should be centered about $\omega_0$. As in the scalar
case \cite{skaar_waagaard}, we also note that in general, the
geometrical phase variation $\theta(x)$ cannot be distinguished from
the phase variation associated with the dc index term $\Delta
n_{\text{dc}}(r,\phi,x)$.

We observe that $\m\sigma$ is real and symmetric, and $\m\kappa$ is
imaginary and symmetric. Moreover, it is not difficult to realize that
$i\m\kappa$ is positive semidefinite.\footnote{The real matrix given
by the elements $\Psi_p\Psi_q$ is clearly positive semidefinite, since
$\sum_{p,q} a_p \Psi_p\Psi_q a_q=(\sum_p\Psi_p a_p)^2\geq 0$ for any
real $a_p$. For a fiber grating $\Delta n_{\text{ac}}(r,\phi,x)\geq 0$
for all $r$ and $\phi$; thus $|\kappa_{pq}(x)|$ adopts the positive
semidefinite property from $\Psi_p\Psi_q$.} Thus $\m\Phi_j$ defined in
\eqref{eq:Tmatrs} is unitary and symmetric, and $-\m\rho_j$ is real
and positive semidefinite. It follows that we can use the
layer-stripping method together with the factorization approach c), as
given in Section \ref{sec:lp}, to identify the coupling sections
$\m\rho_j$ and $\m\Phi_j$ (and therefore the coupling matrices
$\m\kappa$ and $\m\sigma$ as a function of position $x$). Since
$(i\m\Phi_j)^\T(-\m\rho_j)(i\m\Phi_j)=\m\Phi_j^\T\m\rho_j\m\Phi_j$,
the factorization approach gives $-\m\rho_j$ and $i\m\Phi_j$.

For a fiber grating it is usually reasonable to assume that the ac and
dc index modulations can be written in the forms $\Delta
n_\text{ac}(r,\phi,x)=\Delta n(r,\phi)\Delta n_\text{ac}(x)$ and
$\Delta n_\text{dc}(r,\phi,x)=\Delta n(r,\phi)\Delta n_\text{dc}(x)$,
respectively. Here $\Delta n(r,\phi)$ accounts for the transversal
variation of the index modulation profile, and $\Delta n_\text{ac}(x)$
and $\Delta n_\text{dc}(x)$ are the ac and dc modulations as a
function of $x$. As before, we assume that the index modulation and
$n_\text{co}-n_\text{cl}$ are small, yielding
\begin{subequations}
   \label{eq:92}
   \begin{align} 
   \label{eq:92a}
     \m\kappa(x)&=-i\frac{\Delta n_\text{ac}(x)}{2}\m \eta, \\
   \label{eq:92b}
     \m\sigma(x)&=\Delta n_\text{dc}(x)\m \eta
     -\frac{1}{2}\frac{\diff \theta(x)}{\diff x}\m I,
   \end{align}
\end{subequations} 
where $\m\eta$ is independent on $x$. The elements
of $\m\eta$ are
\begin{equation}
\label{eq:93}
 \eta_{pq}=k\int_{A_\infty} \Delta
n(r,\phi)\Psi_p\Psi_q\diff A.
\end{equation} 
When the mode profiles and $\Delta n(r,\phi)$ are
known, this means that the entire coupling matrix $\m\kappa(x)$ is
determined from only a single nonvanishing element. For $\m\sigma$,
two elements are needed (including at least one diagonal
element). Note that in this case, it is indeed possible to distinguish
between the dc index modulation $\Delta n_{\text{dc}}(x)$ and the
geometrical phase variation $\diff\theta(x)/\diff x$ using information contained in $\m\sigma$.

\begin{figure}
    \centering
    \includegraphics{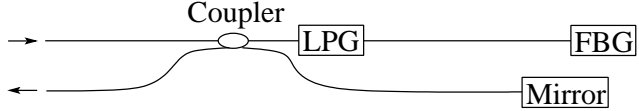}
    \caption{Measurement setup for characterization of multimode gratings.}
    \label{fig:meassetup}
  \end{figure}

For characterization of multimode gratings, measurements
  of the reflection from every mode to every mode are required.
  Performing such measurements is not trivial.  In Ref.
  \cite{rosenhal05:_exper}, an auxiliary long-period
  grating (LPG), i.e, a grating with purely codirectional coupling, is used to
  characterize another interrogated LPG. Fig.
  \ref{fig:meassetup} shows how this method can be adopted to
  characterization of multimode fiber Bragg gratings (FBGs) using optical frequency domain
  reflectometry \cite{foggatt96:_distr}, provided there are no degenerated modes. Light is coupled into
  the fundamental mode of the input fiber and the frequency of the
  highly coherent source is swept. The coupler splits the light
  equally into two fibers. The LPG couples light from the
  fundamental modes into the other modes so that the total optical
  power is distributed between all modes. The light returned by the
  FBG will again propagate through the LPG, and some light from each
  mode will be coupled back into the fundamental mode. The mirror
  reflects only the fundamental mode, and at the coupler the reflected
  light from the mirror interferes with the light in the fundamental
  mode out of the LPG. If the fiber between the LPG and FBG is sufficiently long
  such that the difference in delay between the modes is larger than
  the length of the impulse response of the FBG, the individual
  elements of the reflection matrix will be separable in the
  time-domain.

\section{Numerical example}

A potential application of the multimode layer-stripping method is to
characterize coupling from the core mode to cladding modes in a single
mode fiber. Cladding modes are not bound within the core
of the fiber, but by the cladding/air boundary \cite{Erdogan97}. A
single mode fiber may support as many as 100 cladding modes. The
power in these modes will eventually be lost to the environment. The
core-cladding mode coupling can be seen clearly in the transmission
spectra of strong gratings. For chirped gratings \cite{Ouellette87}
and chirped, sampled gratings \cite{OuelletteKSDE95}, the bandwidth may become larger than
the separation in resonant wavelength between the core-core mode coupling and
the core-cladding mode coupling. Then the core-cladding mode coupling
will interfere with the reflection spectrum associated with the core mode
\cite{finazzi01}. This unwanted coupling is often handled by writing
the grating in fibers with depressed cladding modes
\cite{DongRCCdP97}. There has also been some attempts of taking into
account the core-cladding mode coupling in the design of the grating
\cite{LiNOSR05,ghiringhelli03}. Here, direct scattering is treated
with multiple mode coupling, but the inverse scattering has so far been purely
single-mode. The layer-stripping algorithm described in Section
\ref{sec:lp} can be used for characterization of such coupling
and possibly for design. In contrast to the methods in
\cite{LiNOSR05,ghiringhelli03}, multiple modes can be taken into
account in the inverse scattering part of an iterative design
process. 

A simpler, but nevertheless interesting problem is to characterize
coupling in an optical fiber with a few bound modes. Here, we will
present a numerical experiment simulating a grating in a fiber with
$n_\text{co}=1.452$, $n_\text{cl}=1.437$, and core radius
$r_\text{co}=5\: \mu$m. By solving the eigenvalue equation for a
circular fiber \cite{snyder}, we find that this fiber supports four
modes: LP$_{01}$, LP$_{11}$, LP$_{21}$, and LP$_{02}$ at the design
wavelength $\lambda_0=$ 1.55 $\mu$m. Here, the index $l$ in LP$_{lm}$
means that the transversal field profile can be written in the form
$f_{lm}(r)\cos(l\phi)$. In the further discussion, these modes are
denoted 1 to 4 in the order indicated above. The eigenvalue equation
gives the modal indices $n_1$=1.449, $n_2$=1.444, $n_3$=1.439, and
$n_4$=1.437. We assume that the refractive index is modulated
uniformly in the core of the fiber, but not at all in the cladding.
This is quite realistic since, during fabrication, the fiber usually
is made sensitive to UV exposure only in the core. By evaluating
\eqref{eq:93}, we find that there will be no coupling between modes
with different azimuthal indices $l$:
\begin{equation}
  \label{eq:93c}
  \m\eta=\frac{2\pi}{\lambda_0}
  \matrise{0.957 & 0 & 0 & -0.116\\
              0 & 0.874 & 0 & 0\\
              0 & 0 & 0.707 & 0\\
             -0.116 & 0 & 0 & 0.491}.
\end{equation}
There is no coupling to or from modes 2 and 3; thus the grating
profile can be found by applying a scalar layer-stripping method
separately to the responses associated with these modes. On the other
hand, modes 1 and 4 are coupled, so that the multimode layer-stripping
method must be applied when using the associated responses as a
starting point.

\begin{figure}[htb]
  \includegraphics{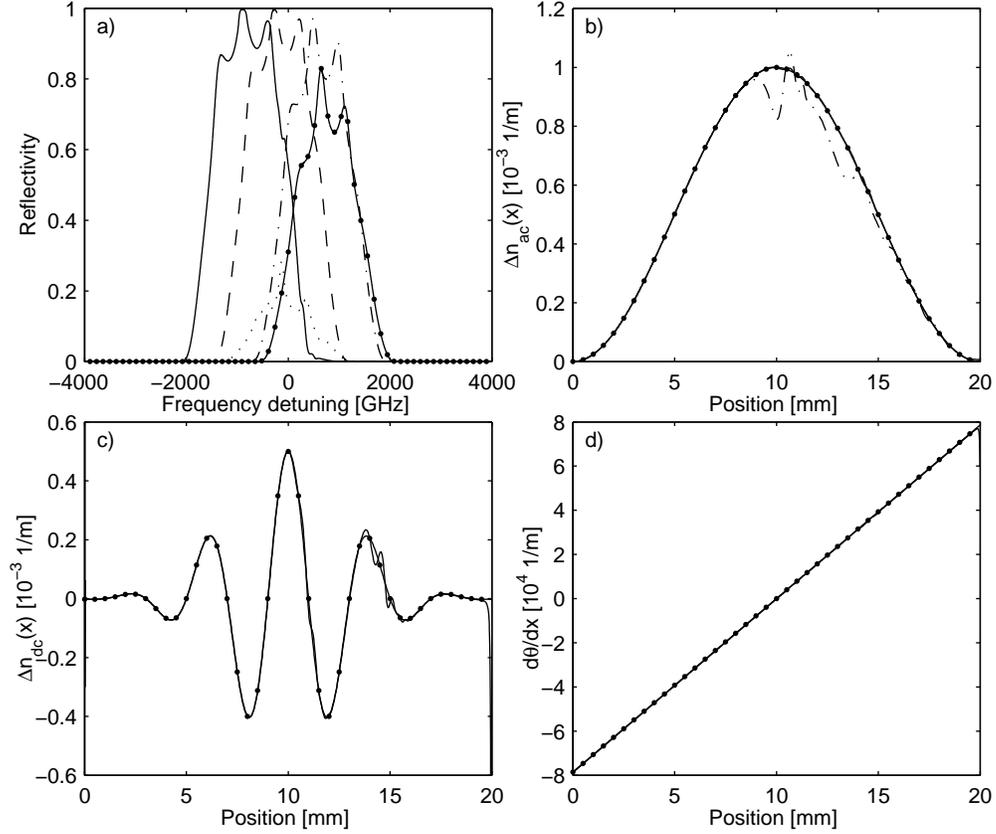}
  \caption{a) Magnitude of the reflection spectrum $|R_{11}|$ (solid
    curve), $|R_{22}|$ (dashed curve),
    $|R_{33}|$ (dashed-dotted curve), $|R_{44}|$ (solid point-marked
    curve), $|R_{14}|=|R_{41}|$ (dotted curve).
    b) Reconstructed longitudinal ac modulation $\Delta
    n_\text{ac}(x)$ (solid curve), actual ac modulation (solid
    point-marked curve) and ac modulation calculated using scalar
    layer-stripping on $R_{11}$ (dashed-dotted curve).
    c) Reconstructed longitudinal dc modulation $\Delta
    n_\text{dc}(x)$ (solid curve), actual dc modulation (solid
    point-marked curve) 
    d) Reconstructed grating phase gradient $\diff\theta/\diff x$
    (solid curve) and actual grating phase gradient (solid
    point-marked curve).}
  \label{fig:numex}
\end{figure}

Defining the nominal mode index $n_0=(n_1+n_4)/2$, the grating period
is set to $\Lambda=\lambda_0/(2n_0)$. The length of the grating is
$L=20$~mm, and $\Delta n_\text{ac}(x)$ has the form of a raised cosine
window with maximum value $\pot{1}{-3}$. Furthermore, $\Delta
n_\text{dc}(x)$ is chosen as a sine-modulated Gaussian window with
full-width-at-half-maximum of 7 mm and a maximum value $\pot{5}{-4}$;
the period of the sine-modulation is 4 mm. The grating is chirped by
varying the grating phase according to
\begin{equation}
  \label{eq:2}
  \frac{\diff\theta}{\diff x}=\pot{\frac{\pi}{8}}{4}\left(x-\frac{L}{2}\right)~\text{m}^{-1}.
\end{equation}
The reflection matrix as a function of frequency detuning is generated
using the piecewise uniform approximation (Section \ref{sec:model})
with $\Delta x=10\:\mu$m, which gives $N$=2000. Zero detuning is taken
to be the frequency $f_0=c/\lambda_0$. Figure \ref{fig:numex}a) shows
the resulting reflection matrix spectrum. The maximum values are
$[|R_{11}|,|R_{22}|,|R_{33}|,|R_{44}|,|R_{14}|]_\text{max}= [99.6,\:
99.6,\: 97.0,\: 83.0,\: 28.3\:]$\%.  Note that the large chirp has
resulted in significant spectral overlap between the different
elements.

The reflection matrix is applied as input to the layer-stripping
method. As the modal indices are similar in magnitude, we use the
discrete algorithm directly, and $\m\Upsilon_j$ is calculated
by taking into account the factor $(n_p+n_q)/(2n_0)$ as discussed in
Section \ref{sec:lp}. Moreover, $\m\kappa(x)$ and $\m\sigma(x)$
is calculated by inverting the expressions for $\m\rho_j$ and
$\m\Phi_j$ in \eqref{eq:Trho1} and \eqref{eq:TPhi1}, respectively.
Figure \ref{fig:numex}b) shows $\Delta n_\text{ac}(x)$ along with its
reconstructed version. The reconstructed $\Delta n_\text{ac}(x)$ is
calculated by a least square fit to \eqref{eq:92a} using the diagonal
elements of the reconstructed $\m\kappa(x)$. We find that the error in
reconstructed profile is less that $\pot{4}{-6}~\text{m}^{-1}$. Also
shown is the ac modulation profile calculated using scalar
layer-stripping on $R_{11}$. Due to the strong coupling between mode 1
and 4, the scalar layer-stripping method does not reconstruct the
profile accurately. Figure \ref{fig:numex}c) and \ref{fig:numex}d)
show that it is possible to separate the dc index variations $\Delta
n_\text{dc}(x)$ from the grating phase gradient $\diff\theta(x)/\diff
x$. The separation is based on a least square fit to \eqref{eq:92b}
using the diagonal elements of $\m\sigma(x)$. The error in
reconstructed $\Delta n_\text{dc}(x)$ is less than
$\pot{6}{-5}~\text{m}^{-1}$, while the error in reconstructed
$\diff\theta(x)/\diff x$ is less than $300~\text{m}^{-1}$. Errors are
mainly due to the finite $\Delta x$ in addition to the fact that the
reflection matrix spectrum of the discretized structure is strictly
nonperiodic (see last paragraph of Section \ref{sec:lp}).

\section{Analogies to 3D inverse scattering}
\label{sec:three-dimens-coupl}

An important inverse scattering problem is the three-dimensional
problem associated with the Schr\"{o}dinger equation
\cite{newton_is3d},
\begin{equation}
  \label{eq:schrodinger}
  \left\{\nabla^2+k^2-V(x,y,z)\right\}\psi(x,y,z;k)=0,
\end{equation} 
where $\psi(x,y,z,k)$ is the wave function and
$V(x,y,z)$ is a smooth and nonnegative potential with compact
support. In particular, solutions to this problem is applicable to
inverse seismic scattering. This problem has been solved using a
generalized Marchenko method in \cite{newton_is3d} and
\cite{rose_is3d}, while layer-stripping solutions are suggested in
\cite{YagleL86} and \cite{Yagle86}. Note the close resemblance between
\eqref{eq:schrodinger} and \eqref{eq:72}, indicating that a similar
method as that in Section \ref{sec:cont-coupl-struct} can be used.

We express the solution as a superposition of the eigenmodes of the
Schr\"{o}dinger equation with $V(x,y,z)=0$. Writing
$\psi(x,y,z;k)=\Psi(y,z;k_y,k_z)\exp(ik_x x)$, these eigenmodes are
given by
\begin{equation}
  \label{eq:85}
  \Psi(y,z;k_y,k_z)=\exp(i(k_yy+k_zz)),
\end{equation} 
where $k_y$ and $k_z$ are the wave numbers in
$y$-direction and $z$-direction, respectively, and
$k^2=k_x^2+k_y^2+k_z^2$.

In a discrete model, the wavenumbers $k_y$ and $k_z$ can for example
be discretized in equal intervals $\Delta k$, such that $k_y=p\Delta
k$ and $k_z=q\Delta k$. In the $yz$-plane, this means that only a
principal range $(-\pi/\Delta k,\pi/\Delta k)$ is considered, and the
fields are extended periodically outside this range. The integers $p$
and $q$ are the modal indices satisfying $p^2+q^2\le (k/\Delta k)^2$
for propagating (not evanescent) modes. The modal field profiles are
written in normalized form $\Psi_{pq}(y,z)=(\Delta
k/2\pi)\Psi(y,z;p\Delta k,q\Delta k)$. The total field $\psi(x,y,z;k)$
is expressed as the superposition
\begin{equation}
  \label{eq:87}
  \psi(x,y,z;k)=\sum_{p,q}
  (b^+_{pq}(x)+b^-_{pq}(x))\Psi_{pq}(y,z),
\end{equation}
 where $b^\pm_{pq}(x)$ includes all $x$-dependence of
the fields, and $\pm$ indicate the sign of $k_x$, i.e, the propagation
direction of the mode.

As in Section \ref{sec:cont-coupl-struct}, we insert \eqref{eq:87}
into \eqref{eq:schrodinger}, multiply by $\Psi^*_{pq}(y,z)$ and
integrate over the principal range of the $yz$-plane. This leads to
the coupled mode equations
\begin{subequations}
  \label{eq:88}
  \begin{align}
    \label{eq:89} \frac{\diff b^+_{pq}(x)}{\diff x}-ik_{x,pq}
    b^+_{pq}(x) &=i\sum_{r,s}\mathcal
    C_{pq,rs}(x)(b^+_{rs}(x)+b^-_{rs}(x)), \\
    \label{eq:90} \frac{\diff b^-_{pq}(x)}{\diff x}+ik_{x,pq}
    b^-_{pq}(x) &=-i\sum_{r,s}\mathcal
    C_{pq,rs}(x)(b^+_{rs}(x)+b^-_{rs}(x)),
  \end{align}
\end{subequations} 
where the coupling coefficients are given by
\begin{equation}
  \label{eq:91}
  \begin{split} \mathcal C_{pq,rs}(x)& = -\frac{1}{2k_x}\int
    \Psi^*_{pq}(y,z)V(x,y,z)\Psi_{rs}(y,z)\diff y\diff z \\ &=
    -\frac{1}{2k_x}\left(\frac{\Delta k}{2\pi}\right)^2\int V(x,y,z)
    \exp\left[i\Delta k((r-p)y+(s-q)z)\right]\diff y\diff z,
  \end{split}
\end{equation} 
and $k_{x,pq}=[k^2-(\Delta k)^2(p^2+q^2)]^{1/2}$. We
restrict ourselves to the situation where $V(x,y,z)$ is known to be
quasi-periodic along the $x$-direction. Then an expansion of $\mathcal
C_{pq,rs}(x)$ as in \eqref{eq:quasisin} together with the
transformation \eqref{eq:69} can be used, resulting in the exact same
problem as that described in Section \ref{sec:cont-coupl-struct}. Thus
the layer-stripping method in Section \ref{sec:lp} can be applied. The
required input data is the reflection into all plane waves upon
excitation of the different plane waves onto the plane $x=0$. The
scattering potential $V(x,y,z)$ is found from the inverse of
\eqref{eq:91}.

There are two complications. First, in order to use the factorization
methods developed in Section \ref{sec:lp}, we must ensure that
reciprocity implies symmetric scattering matrices. This is guaranteed
when the mode profiles can be written real. Thus we define real mode
fields by the transformation
\begin{equation}
  \label{eq:mtransf} \matrise{{\m\Psi}_{++}\\ {\m\Psi}_{-+}\\
    {\m\Psi}_{+-}\\ {\m\Psi}_{--}} \to \mathbf{M}
    \matrise{{\m\Psi}_{++}\\ {\m\Psi}_{-+}\\ {\m\Psi}_{+-}\\
    {\m\Psi}_{--}}, \quad \mathbf{M}=\frac{1}{2}\matrise{
\m I & \m I & \m I & \m I \\
-i\m I & +i\m I & -i\m I & +i\m I \\ i\m I & -i\m I & +i\m I & +i\m I
-\\ \m I & \m I & \m I & -\m I }.
\end{equation}
 Here, ${\m\Psi}_{++}$ denotes a column vector
containing the modal field amplitudes $\Psi_{pq}$ with positive $p$
and $q$; ${\m\Psi}_{-+}$ denotes a column vector containing the modal
field amplitudes with negative $p$ and positive $q$, and so forth. The
dimension of the identity matrices in the blocks of $\mathbf{M}$
corresponds to the dimension of ${\m\Psi}_{++}$. If
$\mathbf{\mathcal{C}}$ denotes the matrix formed by the elements
$\mathcal C_{pq,rs}$, the coupling matrix transforms $\mathbf{\mathcal
C}\to\mathbf{M}^*\mathbf{\mathcal{C}}\mathbf{M}^\T$. Inspection of
\eqref{eq:91} shows that the transformed $-\mathbf{\mathcal{C}}$ is
real and positive semidefinite (recall that $V(x,y,z)\geq 0$); thus
enabling the factorization method in Section \ref{sec:lp}.

Second, the causality argument of the layer-stripping method does only
work when the coupling matrix $\mathbf{\mathcal C}$ is independent on
frequency. Eq. \eqref{eq:91} shows that this condition can only be
justified when the relevant frequency band is narrow. Therefore the
structure must, in addition to be quasi-periodic along the
$x$-direction, vary slowly along the transversal direction. The
variation must be sufficiently slow such that the modes with
$(p^2+q^2)\Delta k^2\ll k^2$ contain sufficient information about the
transversal dependence, and the other modes may be neglected.

\section{Conclusion}
 A layer-stripping method for the inverse
scattering of multi-mode structures has been proposed. Ambiguities
related to factorization of each layer's response into codirectional
and contradirectional coupling have been discussed. When there are no
codirectional coupling, the ambiguities disappear. Also, when the
structure to be reconstructed is smooth, there are important cases
with simultaneous co- and contradirectional coupling that can be
reconstructed uniquely, provided the reflector eigenvalues are nonzero
and nondegenerate. Applications to quasi-periodical structures, and
analogies to multidimensional inverse scattering have been discussed.

\begin{appendix}

\section{Matrix factorizations}
\label{sec:matrprop}

\subsection{Takagi factorization of complex symmetric matrices}
\label{sec:svdsymm} Any complex symmtric matrix $\m\Upsilon$ can be
written
\begin{equation}
\label{eq:symmsvd} 
\m \Upsilon =\m U^\T \m\Sigma \m U,
\end{equation}
where $\m U$ is unitary and $\m\Sigma$ is
diagonal and nonnegative (See e.g. \cite{horn_johnson}, Chapter
4.4). Eq. \eqref{eq:symmsvd} is called Takagi factorization.

A constructive proof, suitable for implementation, can be given as
follows: Singular value decomposition yields
\begin{equation}
 \m\Upsilon=\m V_1 \m\Sigma \m V_2,
\end{equation} 
where $\m V_{1,2}$ are unitary, and $\m\Sigma$ is
diagonal and nonnegative. Using $\m\Upsilon=\m\Upsilon^\T$ and
$(\m\Upsilon\m\Upsilon^\dagger)^\T=\m\Upsilon^\dagger\m\Upsilon$ we
find that $\m W \m \Sigma = \m \Sigma \m W^\T=\m\Sigma \m W$, where
$\m W\equiv \m V_2^*\m V_1$. Thus, provided $\m\Upsilon$ is
nonsingular, $\m W$ is symmetric. Then $\sqrt{\m W}$ can be chosen
such that it commutes with $\m \Sigma$ and is symmetric, and we obtain
$\m\Upsilon=\m V_2^\T \m W \m\Sigma \m V_2 = (\sqrt{\m W}\m V_2)^\T
\m\Sigma \sqrt{\m W}\m V_2$, or
\begin{equation}
\m \Upsilon =\m U^\T \m\Sigma \m U,
\end{equation} 
where $\m U\equiv \sqrt{\m W}\m V_2$ is unitary and
$\m\Sigma$ is diagonal and positive.

If $\m\Upsilon$ is singular, we write
\begin{equation}
  \label{eq:23} 
  \m \Sigma=\matrise{\m \Sigma' & \m 0 \\ \m 0 & \m 0}
  \qquad \text{and} \qquad 
  \m W=\matrise{\m W_{11} & \m W_{12} \\ \m W_{21} & \m W_{22}},
\end{equation} 
where we have arranged $\m \Sigma$ so that the zero
singular values are the last ones, $\m \Sigma'$ is a diagonal matrix
with the nonzero singular values, and $\m W_{11}$ has the same
dimension as $\m \Sigma'$. We now find $\m \Sigma' \m W_{11}=\m
W_{11}\m \Sigma'$, $\m W_{12}=\m W_{21}=\m 0$, and $\m W_{11}=\m
W_{11}^\T$. The commutation relations do not provide any information
on $\m W_{22}$. Choose $\sqrt{\m W}$ such that
\begin{equation} 
\sqrt{\m W}=\matrise{\sqrt{\m W_{11}} & \m 0 \\ \m 0
  & \sqrt{\m W_{22}}},
\end{equation}
where $\sqrt{\m W_{11}}$ is symmetric and $\sqrt{\m
W_{11}}$ and $\m\Sigma'$ commute. Write $\m \Upsilon=\m U_1^\T \m
\Sigma \m U_2$, with
\begin{align}
  \label{eq:25} 
  \m U_1&=\sqrt{\m W}^\T\m V_2=\matrise{\m U' \\ \m  U''_1}\\ 
  \m U_2&=\sqrt{\m W}\m V_2=\matrise{\m U' \\ \m U''_2}.
\end{align} 
The matrices $\m U''_1$ and $\m U''_2$ are the rows of $\m
U_1$ and $\m U_2$ that correspond to the zero singular values, and
they do not give any contribution to $\m \Upsilon$. We may therefore
replace the rows $\m U''_1$ by $\m U''_2$, which gives $\m U_1=\m
U_2=\m U$.

The matrix $\m\Sigma$ is unique up to reordering of the singular
values. When the order of the singular values is established, $\m U$
is unique up to the replacement $ \m J\m U \to \m U$, where $\m J$ is
a unitary matrix satisfying $(\m J \m U)^\T \m\Sigma \m J \m U =\m
U^\T \m\Sigma \m U$. This leads to $\m J^\T \m \Sigma \m J=\m \Sigma$.
Assuming the singular values are sorted in, say, descending order, we
find that $\m J$ is a unitary block-diagonal matrix, where each block
has a dimension equal to the number of corresponding repeated singular
values. For zero singular values, the corresponding block in $\m J$ is
an arbitrary unitary matrix. For repeated non-zero singular values,
the corresponding block in $\m J$ is real. For a distinct, non-zero
singular value, the corresponding block of $\m J$ is either 1 or $-1$.

\subsection{Factorization of a unitary matrix into a symmetric matrix
and an orthogonal matrix}
\label{sec:UPQfact}

A unitary matrix $\m U$ can be factorized into $\m U=\m P \m\Phi$,
where $\m P$ is a real unitary matrix (orthogonal matrix) and $\m\Phi$
is a symmetric unitary matrix (See e.g. \cite{horn_johnson}, Chapter
3.4). A constructive proof, suitable for implementation, can be given
as follows. First we note that the symmetric unitary matrix $\m\Phi$
can be factorized into $\m\Phi=\m P_1 \m D \m P_1^\T$, where $\m D$ is
a diagonal unitary matrix and $\m P_1$ is a real unitary matrix (a
simple, constructive proof for this particular spectral decomposition
is given in \cite{horn_johnson}, Chapter 4.4). Thus, an equivalent
problem is to show that
\begin{equation}
  \label{eq:22}
  \m U=\m P_2 \m D \m P_1^\T,
\end{equation} 
where $\m P_2=\m P \m P_1$. The decomposition in
\eqref{eq:22} is very similar to singular value decomposition of real
matrices, except that $\m D$ may have complex elements.

The matrix $\m U^\T \m U$ is unitary and symmetric; thus we can write
\begin{equation}
  \label{eq:30} 
  \m U^\T \m U=\m P_1 \m \Lambda \m P_1^\T,
\end{equation} 
where $\m P_1$ is a real unitary matrix and $\m
\Lambda$ is a diagonal unitary matrix. Define
\begin{equation}
  \label{eq:55} 
  \m P_2=\m U \m P_1 \m D^*,
\end{equation}
 where the diagonal matrix $\m D$ is a solution to $\m
D^2=\m \Lambda$. The matrix $\m P_2$ is unitary since it is produced
by multiplication of unitary matrices, thus $\m P_2^* \m P_2^\T=\m I$.
The matrix is also real since
\begin{equation}
\label{eq:56}
\m P_2^\T \m P_2 =\m D^* \m P_1^\T \m U^\T
  \m U \m P_1 \m D^* = \m D^* \m P_1^\T \m P_1 \m D^2 \m
  P_1^\T \m P_1 \m D^*=\m I,
\end{equation} 
which gives $\m P_2=(\m P_2^* \m P_2^\T)\m P_2=\m P_2^* (\m P_2^\T\m
P_2)=\m P_2^*$.

From \eqref{eq:55} we therefore conclude that the
decomposition \eqref{eq:22}, with real unitary $\m P_1$ and $\m P_2$
and diagonal $\m D$, is always possible. It follows that any unitary
matrix can be written $\m U=\m P \m\Phi$, where $\m P$ is real and
unitary, and $\m\Phi$ is symmetric and unitary. Note that any global
phase of $\m P$ can instead be assigned to $\m\Phi$, so without loss
of generality we can assume that $\m P$ is special ($\det\m P=1$ and
$\det\m\Phi=\det\m U$).

Since $\m D$ is calculated from $\m D^2=\m \Lambda$, the sign of its
elements are arbitrary. The ambiguities when determining $\m P_1$ in
\eqref{eq:30} give rise to ambiguities in $\m P$ and $\m\Phi$. The
possible $\m P$ and $\m\Phi$ can be expressed as $\m P=\m U\m P_1\m
J\m D^*\m J^\T\m P_1^\T$ and $\m\Phi=\m P_1\m J\m D\m J^T\m P_1^\T$
for a real unitary $\m J$ that commutes with $\m D^2$. Here $\m P_1$
is fixed. If the signs of the elements of $\m D$ are known to be such
that any equal elements of $\m D^2$ correspond to equal elements of
$\m D$, then $\m J$ commutes with $\m D$ and can be ignored.

\section{Linear, reciprocal and lossless components}
\label{sec:gencomp}

\begin{figure}[h] 
  \centering
  \includegraphics[width=2.5cm]{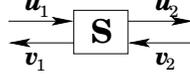}
  \caption{A linear component with $P$ input and $P$ output modes on
    each side.}
  \label{fig:comp}
\end{figure}

Consider a linear component with $P$ input and $P$ output modes on the
left-hand side, and also $P$ input and $P$ output modes on the
right-hand side, see Fig. \ref{fig:comp}.
The component is completely characterized by the
$2P\times 2P$ dimensional scattering matrix $\mathbf{S}$ which relates
the input and output fields:
\begin{equation}
  \label{eq:21} 
  \matrise{\m v_1\\\m u_2}=\mathbf S \matrise{\m u_1\\\m v_2} 
  =\matrise{\m S_{11} & \m S_{12}\\ \m S_{21} & \m S_{22}}
  \matrise{\m u_1\\\m v_2}.
\end{equation} 
The field vectors that propagate to the right and left
are denoted $\m u$ and $\m v$, respectively, and the subscripts 1 and
2 indicate the left- and right-hand side of the component. The
scattering matrix is a block matrix; the blocks $\m S_{11}$ and $\m
S_{22}$ being the reflection from the left and right side of the
device, respectively, and $\m S_{21}$ and $\m S_{12}$ the transmission
through the device from the left and right, respectively. These blocks
have the dimension $P\times P$.

There exists a similar relation, a transfer matrix relation, that
connects the fields on the left-hand side to the fields on the
right-hand side:
\begin{equation}
  \label{eq:58} 
  \matrise{\m u_2\\\m v_2}=\mathbf T \matrise{\m u_1\\\m  v_1}.
\end{equation} 
Comparing \eqref{eq:21} and \eqref{eq:58} we find the
blocks of $\mathbf{T}$:
\begin{equation}
  \label{eq:11}
  \mathbf T=\matrise{\m S_{21}-\m S_{22}\m S_{12}^{-1}\m S_{11} & \m
    S_{22}\m S_{12}^{-1}\\ -\m S_{12}^{-1}\m S_{11} & \m S_{12}^{-1}}.
\end{equation} 
To describe a device with a transfer matrix, $\m
S_{12}$ must be invertible, that is the transmission from the right
cannot be zero for any input field vector. Thus, ideal mirrors, for
example, cannot be described by a transfer matrix.

Provided the mode profiles can be written real, reciprocity means that
the scattering matrix is symmetric \cite{pozar,haus}, i.e.,
\begin{subequations}
  \label{eq:13}
  \begin{align} 
    \m S_{11}&=\m S_{11}^\T \label{eq:14}\\ 
    \m S_{22}&=\m S_{22}^\T \label{eq:15}\\ 
    \m S_{21}&=\m S_{12}^\T \label{eq:16}.
  \end{align}
\end{subequations} 
Moreover, the lossless condition is expressed as
the unitarity condition $\mathbf S^\dagger \mathbf S=\mathbf I$:
\begin{subequations}
  \label{eq:26}
  \begin{align} 
    \m S_{11}^\dagger \m S_{11}+\m S_{21}^\dagger \m S_{21}&=\m I
    \label{eq:27}\\ 
    \m S_{12}^\dagger \m S_{12}+\m  S_{22}^\dagger \m S_{22}&=\m I
    \label{eq:28}\\ 
    \m S_{12}^\dagger \m S_{11}+\m S_{22}^\dagger \m S_{21}&=\m 0 \label{eq:29}.
  \end{align}
\end{subequations}
With \eqref{eq:13} in mind, we introduce Takagi
factorization of $\m S_{11}$ and $-\m S_{22}$ (see Appendix
\ref{sec:svdsymm}):
\begin{subequations}
  \label{eq:svdrt}
  \begin{align} 
    \m S_{11}&=\m\Phi_\text{l}^\T \m\rho\m\Phi_\text{l}
    \label{eq:14svd}\\ 
    \m S_{22}&=\m\Phi_\text{r} (-\m\rho')\m\Phi_\text{r}^\T
    \label{eq:15svd}\\ 
    \m S_{21}&=\m\Phi_\text{r} \m t'\m\Phi_\text{l} 
    \label{eq:16svd}.
  \end{align}
\end{subequations} 
Here, $\m\Phi_\text{l}$ and $\m\Phi_\text{r}$ are
unitary matrices, $\m\rho$ and $\m\rho'$ are diagonal and nonnegative,
and $\m t'=\m\Phi_\text{r}^\dagger\m
S_{21}\m\Phi_\text{l}^\dagger$. By substituting into \eqref{eq:26} and
using \eqref{eq:13} we obtain
\begin{subequations}
  \label{eq:26m}
  \begin{align} 
    \m t'^\dagger \m t'=\m I-\m\rho^2\label{eq:14m}\\ 
    \m t' \m t'^\dagger=\m I-\m\rho'^2\label{eq:15m}\\ 
    \m\rho'=\m t'\m\rho \m t'^{*-1}\label{eq:16m}.
  \end{align}
\end{subequations} 
Introducing the singular value decomposition $\m t'=\m U'\m t\m V'$, 
we obtain from \eqref{eq:14m} that $\m t^2=\m V'(\m I-\m\rho^2)\m
V'^\dagger$, which means $\m t=\m V'\sqrt{\m I-\m\rho^2}\m
V'^\dagger$.  Backsubstitution shows that $\m t'$ can be written $\m
t'=\m U\sqrt{\m I-\m\rho^2}$ for a unitary $\m U$; thus \eqref{eq:16m}
reduces to $\m\rho'=\m U\m\rho\m U^\T$. With these properties, it is
straightforward to show that \eqref{eq:svdrt} can be written
\begin{subequations}
  \label{eq:svdrt2}
  \begin{align} 
    \m S_{11}&=\m\Phi_\text{l}^\T \m\rho\m\Phi_\text{l}
    \label{eq:14svd2}\\ 
    \m S_{22}&=\m\Phi_\text{r} (-\m\rho)\m\Phi_\text{r}^\T
    \label{eq:15svd2} \\ 
    \m S_{21}&= \m S_{12}^\T=\m\Phi_\text{r} \m t\m\Phi_\text{l}
    \label{eq:16svd2},
  \end{align}
\end{subequations} 
where $\m U$ has been absorbed into
$\m\Phi_\text{r}$, $\m\Phi_\text{r}\m U \rightarrow\m\Phi_\text{r}$,
and
\begin{equation}
  \label{eq:tdefr} 
  \m t=\sqrt{\m I-\m\rho^2}.
\end{equation} 
Note that \eqref{eq:26m} implies that $\|\m\rho\|\leq 1$.

Eq. \eqref{eq:svdrt2} and \eqref{eq:tdefr} can be interpreted as
follows: The component can be viewed as a discrete reflector
sandwiched between two unitary transmission sections. The discrete
reflector provides coupling between equal modes that propagate in
opposite directions, and the unitary sections provide coupling between
different modes in the same direction.  For the discrete reflector,
the reflection response from the left and right is $\m\rho$ and
$-\m\rho$, respectively, and the transmission is $\m t$. For the two
unitary sections, there are no reflections, and the transmission
responses from the left are $\m\Phi_\text{l}$ and $\m\Phi_\text{r}$,
while the transmission responses from the right are
$\m\Phi_\text{l}^\T$ and $\m\Phi_\text{r}^\T$. Note that this
interpretation is consistent with the reciprocity and lossless
conditions \eqref{eq:13} and \eqref{eq:26}, for each of the three
sections separately. By inspection, we find that
  \eqref{eq:svdrt2} is invariant if $\m P\m\rho\m
  P^\T\rightarrow\m\rho$, $\m P\m t\m P^\T\rightarrow\m t$, $\m
  P\m\Phi_\text{l}\rightarrow \m\Phi_\text{l}$, and $\m\Phi_\text{r}\m
  P^\T\rightarrow \m\Phi_\text{r}$ where $\m P$ is a real unitary
  matrix. Here $\m P$ represents an arbitrary rotation of the
  eigenaxes of the reflector ($\m\rho$ and $\m t$ are now real and
  positive semidefinite).

Using \eqref{eq:svdrt2}, the transfer matrix \eqref{eq:11} can be
written
\begin{equation}
\label{eq:49} 
\mathbf T=\matrise{\m A^* & \m B^*\\ \m B & \m A},
\end{equation} 
where the blocks $\m A=\m\Phi_r^*\m t^{-1}\m\Phi_l^*$
and $\m B=-\m\Phi_r^*\m t^{-1}\m\rho\m\Phi_l$ satisfy
\begin{subequations}
\label{eq:50}
\begin{align} 
  \m A^\dagger\m A - \m B^\T\m B^*&=\m I\label{eq:51}\\ 
  \m A \m B^\T - \m B \m A^\T &=\m 0 \label{eq:52}\\ 
  \m A^\T \m B^* - \m B^\dagger \m A &=\m 0. \label{eq:53}
\end{align}
\end{subequations}

\end{appendix}

\bibliographystyle{siam} 

\bibliography{fbgdesign,inversescattering}

\end{document}